\documentstyle[12pt,a4,epsf]{article}
\begin{document}
\title{OVERVIEW FROM LATTICE QCD}
\author{Gunnar S.\ Bali\thanks{Plenary talk presented at
``Nuclear and Particle physics with CEBAF at Jefferson Lab'',
Dubrovnik, November 3--10, 1998.}\\
{\small\em Institut f\"ur Physik, Humboldt-Universit\"at zu Berlin,}\\
{\small\em Invalidenstra\ss e 110, 10115 Berlin, Germany}}
\maketitle
\begin{abstract}
I review recent Lattice results. In particular,
the confinement mechanism and string breaking, glueballs and hybrid mesons
as well as light hadron spectroscopy are discussed.
\end{abstract}
\section{Introduction}
The lattice approach to QCD facilitates the numerical evaluation
of expectation values without recourse to perturbative techniques.
Although the lattice formulation
is almost as old as QCD itself and first simulations
of the path integral have been performed as early as in
1979~\cite{creutz},
only recently computers have become powerful enough to allow
for a determination of
the infinite volume continuum
light hadron spectrum in the quenched approximation to QCD within
uncertainties of a few per cent. To this accuracy the quenched
spectrum deviates from experiment.
Some collaborations have started to systematically explore 
QCD with two flavours of light
sea quarks and the first precision results
indeed indicate such differences.

Lattice QCD is a {\em first principles} approach;
no parameters apart from those that are inherent to
QCD, i.e.\ a strong coupling constant at a certain scale
and $n$ quark masses, have to be introduced. In order to fit
these $n+1$ parameters, $n+1$ low energy quantities  are matched
to their experimental values: the lattice spacing $a(g,m_i)$,
that results from given values of the bare lattice coupling
$g$ and (in un-quenched QCD) quark masses $m_i$, can be obtained by fixing
$m_{\rho}$ as determined on the Lattice
to the experimental value.
The lattice parameters that correspond to physical
$m_u\approx m_d$ can then be obtained by adjusting $m_{\pi}/m_{\rho}$;
the right $m_s$ can be reproduced
by adjusting $m_K/m_{\rho}$ or $m_\phi/m_\rho$
to experiment etc.. Once the scale and quark masses
have been set, everything else becomes a prediction. Due to the
evaluation of path integrals by use of a stochastic process,
lattice predictions carry statistical errors which can in principle
be made arbitrarily small by increasing the statistics, i.e.\ the
amount of computer time spent. In this sense, it is an exact approach.

Lattice results have in general to be extrapolated to the
(continuum) limit $a\rightarrow 0$ at fixed physical volume.
The functional form of this extrapolation is theoretically well
understood and under control. This claim is substantiated by the fact
that simulations with different
lattice discretisations of the continuum QCD action yield compatible
results after the continuum extrapolation.
For high energies, an overlap between certain quenched Lattice computations
and perturbative QCD has been confirmed too~\cite{alpha}, excluding the
possibility of fixed points of the $\beta$-function at finite
values of the coupling, other than $g=0$.
After taking the continuum limit an infinite volume
extrapolation is performed. Results on hadron masses from
quenched evaluations
on lattices with spatial extent $La>2$~fm
are virtually indistinguishable from the infinite volume limit
within typical statistical errors in most cases.
However, for QCD with sea quarks
the available information is not yet sufficient for definite conclusions,
in particular as one might expect a substantial dependence of
the on-set of finite size effects on the sea quark mass(es).

The effective infinite volume limit
of realistically light pions cannot be realised
at a reasonable computational cost, neither in quenched
nor in full QCD.
Therefore, in practice another extrapolation, in the quark mass, is
required. This extrapolation to the correct light quark mass limit
is theoretically less well under control than those to
the continuum and infinite volume limits.
The parametrisations used are in general
motivated by chiral perturbation theory
and the theoretical uncertainties
are the dominant source of error in latest state-of-the-art spectrum
calculations.

Many important questions are posed in low-energy QCD:
is the same set of fundamental parameters (QCD coupling and quark masses)
that describes for instance the hadron spectrum consistent with
high energy QCD or is there
place for new physics? Are all hadronic states correctly
classified by the na\"\i{}ve quark model or do glueballs, hybrid states
and molecules play a r\^o{}le?
At what temperatures/densities does the transition to
a quark-gluon plasma occur? What are the experimental signatures
of quark-gluon matter? Can we solve nuclear physics on the quark and
gluon level?  Clearly, complex systems like
iron nuclei are unlikely ever to be solved from {\em first principles} alone
but {\em modelling} and certain {\em approximations} will always be required.

It is desirable to test model assumptions,
to gain control over
approximations and, eventually, to derive low-energy effective
Lagrangians from QCD. Lattice simulations are a very
promising tool for this purpose and in the first part of this article
I will try to give a flavour of such more theoretically motivated
studies before reviewing recent results on glueballs and exotic
hybrid mesons as well as discussing the light hadron spectrum.

\section{The confinement scenario}
Two prominent features of QCD,
confinement of colour sources and spontaneous breaking of chiral symmetry,
are both lacking a proof.
They appear to be related, however: in the low temperature
phase of broken chiral symmetry, colour sources are effectively
confined. Clearly, an understanding of what is going on should help
us in developing the methods required to tackle a huge class of
non-perturbative problems.

It is worthwhile to consider the simpler pure
$SU(N)$ gauge theory. In this case
{\em confinement} can be rigorously defined since
the Polyakov line is an order parameter 
for the de-confinement phase transition
that is related to spontaneously breaking a global $Z_N$ symmetry.
I will present some results that have been obtained in the
computationally cheaper $SU(2)$
gauge theory whose spectrum shares most qualitative
features with that of $SU(3)$.

In the past decades, many explanations of the confinement mechanism
have been proposed, most of which share the feature that topological
excitations of the vacuum
play a major r\^ole. These pictures include, among others, the
dual superconductor picture of confinement~\cite{hooft,hooft2}
and the centre vortex
model~\cite{ambjorn}.
Depending on the underlying scenario, the excitations
giving rise to confinement are thought to be
magnetic monopoles, instantons, dyons,
centre vortices, etc.. Different ideas are not necessarily
exclusive. For instance,
all mentioned excitations are
found to be correlated with each other
in numerical as well as in some analytical studies, such that at present
it seems to be rather
a matter of personal preference which one to consider
as more {\em fundamental}.

\begin{figure}
\vskip -.5truecm
\epsfxsize=12.5truecm
\centerline{\epsfbox{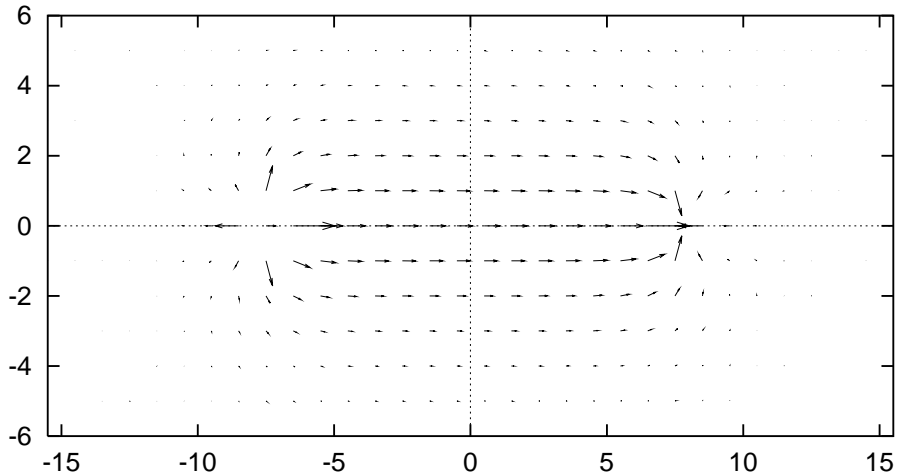}}
\vskip -.5truecm
\caption{Electric field distribution between two static $SU(2)$
  sources
in the MA projection. Everything is in lattice units $a\approx
0.081$~fm.
The sources are located at the coordinates $(-7.5,0)$ and
$(7.5,0)$.} 
\label{fig1}
\end{figure}

Recently, the centre vortex model has enjoyed renewed
attention~\cite{greensite}. In this picture,
excitations that can be classified in accord with
the centre group provide the disorder required to
produce an area law of the Wegner-Wilson loop and, therefore,
confinement. One striking
feature is that --- unlike monopole currents --- centre vortices
form gauge invariant
two-dimensional objects, such that in four space-time dimensions,
a linking number between a Wegner-Wilson loop and centre vortices can
unambiguously be defined, providing a geometric interpretation of
the confinement mechanism~\cite{poli}.

I will restrict myself to discussing the 
superconductor picture which is based on the concept of electro-magnetic
duality after an Abelian gauge projection and has originally been
proposed by 't~Hooft and Mandelstam~\cite{hooft}.
The QCD vacuum is thought to behave analogously
to an electrodynamic superconductor
but with the r\^oles of electric and magnetic fields being interchanged:
a condensate of magnetic monopoles expels electric fields
from the vacuum. If one now puts electric charge and anti-charge
into this medium, the electric flux that forms between them
will be squeezed into a thin, eventually
string-like, Abrikosov-Nielsen-Oleson vortex which
results in linear confinement.

In all quantum field theories in which confinement has been proven,
namely in compact $U(1)$ gauge theory,
the Georgi-Glashow model and SUSY Yang-Mills
theories, this scenario is indeed realised.
However, before one can apply this simple
picture to QCD or $SU(N)$ chromodynamics
one has to identify the relevant dynamical variables: it is not straight
forward to generalise the electro-magnetic duality of a $U(1)$ gauge theory
to $SU(N)$ where gluons carry colour charges. How can one define
electric fields and dual fields in a gauge invariant way?

In the Georgi-Glashow model, the $SO(3)$ gauge symmetry is broken down to
a residual $U(1)$ symmetry as the vacuum expectation value of
the Higgs field becomes finite. It is currently unknown whether QCD
provides a similar mechanism
and various reductions of the SU(N)
symmetry have been conjectured.
In this spirit,
it has been proposed~\cite{hooft2}
to identify the monopoles in a $U(1)^{N-1}$ Cartan subgroup
of $SU(N)$ gauge theory after gauge fixing with respect to the
off-diagonal $SU(N)/U(1)^{N-1}$ degrees of freedom.
After such an Abelian gauge fixing QCD
can be regarded as a theory of interacting
photons, monopoles and matter fields (i.e.\ off-diagonal gluons and quarks).
One might assume that the off-diagonal
gluons do not affect long range interactions.
This conjecture is known as {\em Abelian dominance}~\cite{ezawa}.
Abelian as well as monopole dominance are
qualitatively realised in Lattice studies of
$SU(2)$ gauge theory~\cite{balborn} in maximally Abelian (MA) gauge projection,
which 
appears to be the most suitable gauge fixing condition.

\begin{figure}
\vskip -.5truecm
\epsfxsize=11.0truecm
\centerline{\epsfbox{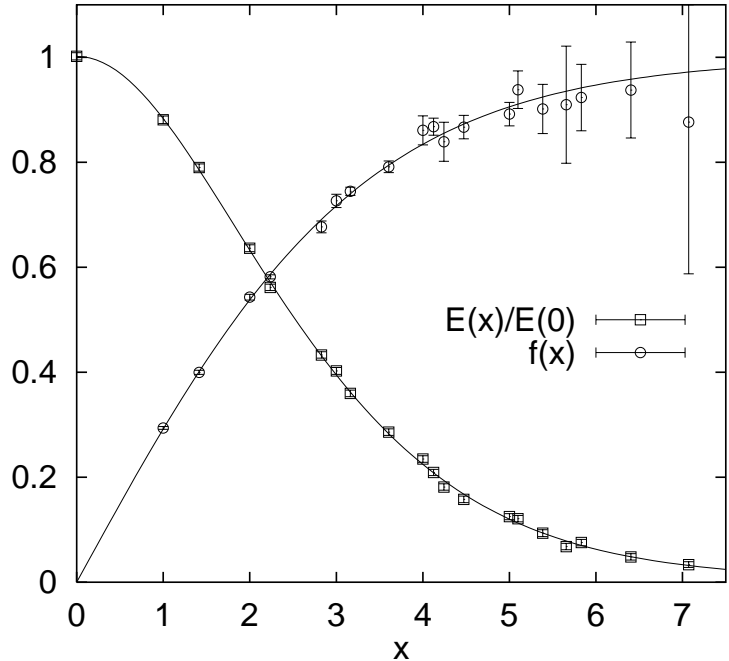}}
\vskip -.5truecm
\caption{Electric field and GL wave function density in the
centre plane between the sources.} 
\label{fig2}
\end{figure}
In Figure~\ref{fig1}, I display the electric field distribution between
SU(2) quarks, separated by a distance $r=15a\approx 1.2$~fm,
that has been obtained within the MA gauge projection.
Everything is measured in lattice units $a\approx 0.081$~fm
where the physical 
scale derived from
the value $\sqrt{\kappa}=440$~MeV for the string tension
is intended to serve as
a guide to what one might expect in ``real'' QCD.
Indeed an elongated Abrikosov-Nielsen-Oleson vortex forms between the charges.
In Fig.~\ref{fig2}, I display a cross section through the centre plane
of this vortex. While the electric field strength decreases with the
distance from the core, the modulus of the dual Ginzburg-Landau (GL)
wave function,
$f$, i.e.\ the density of superconducting magnetic monopoles decreases
towards the centre of the vortex where superconductivity breaks down.
In this study
the values $\lambda=0.15(2)$~fm and $\xi=0.25(3)$~fm have been
obtained
for penetration depth and GL coherence length, respectively.
The ratio $\lambda/\xi=0.59(13)<1/\sqrt{2}$ corresponds to a
(weak) type I superconductor, i.e.\
QCD flux tubes appear to attract each other.
For details I refer the reader to Ref.~\cite{bali}

\section{String breaking}
\begin{figure}
\vskip -.5truecm
\epsfxsize=11truecm
\centerline{\epsfbox{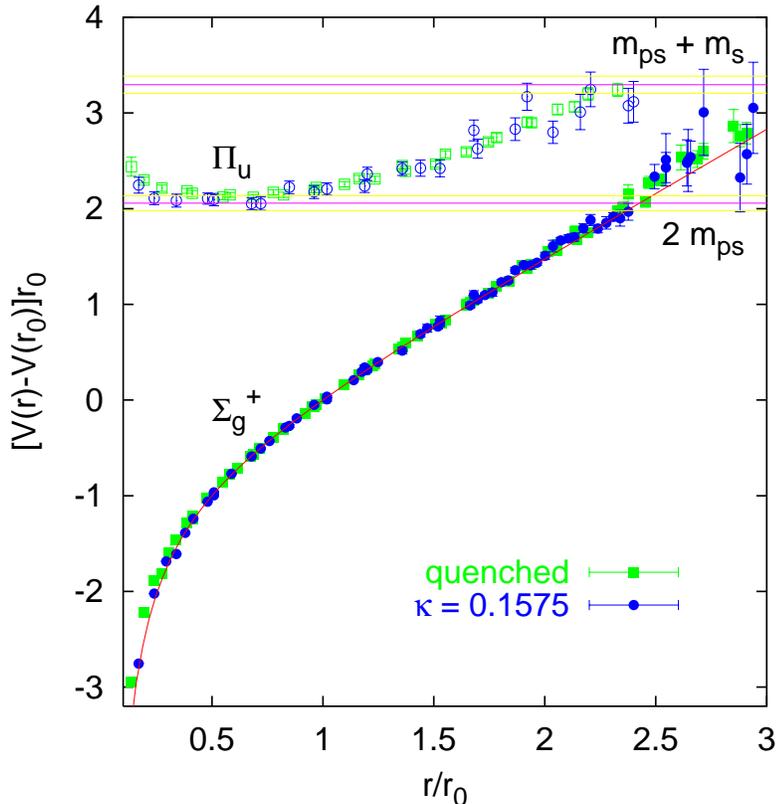}}
\vskip -.5truecm
\caption{The QCD static $\Sigma_g^+$ and $\Pi_u$ potentials, together
with masses of pairs of static-light mesons.} 
\label{fig3}
\end{figure}

In the pure gauge theory results presented above,
the energy stored in the vortex between charges increases in proportion
to their distance {\em ad infinitum}. In full QCD with sea quarks,
however, the string will break into two parts
as soon as this energy exceeds the energy required to create
a quark-antiquark pair from the vacuum:
inter-quark forces at large separation will be completely
screened by sea quarks and excited $\Upsilon$ states can decay into
a $B\overline{B}$ meson pair.
In Fig.~\ref{fig3}, I display a recent comparison between
the quenched and $n_f=2$ static potential by the $T\chi L$ collaboration
at a sea quark mass $m_{ud}\approx m_s/3$~\cite{tcl}.
Estimates of masses of pairs of static-light mesons
into which the static heavy-heavy systems can decay are also included
into the figure.
The potentials have been matched to each other at a
distance $r=r_0 \approx 0.5$~fm.
In presence of sea quarks anti-screening is weakened and, therefore,
starting from the same infra red value, the effective QCD
coupling runs slower towards the $\alpha_s=0$ ultra violet limit.
This effect explains why at small $r$ the unquenched data points
are somewhat below their quenched counterparts: the effective Coulomb
force remains stronger.
Around $r=1.2$~fm, the un-quenched potential is expected to become
unstable. However, the data are not yet precise enough to resolve this
effect.

Motivated by such QCD simulations, the dynamics of string breaking
has recently been analysed in some toy models~\cite{sommer}.
First results on
interactions between two $B$ mesons in quenched QCD
have been reported by
Pennanen and Michael~\cite{petrus}.

\section{Glueballs and quark-gluon hybrid states}
In Fig.~\ref{fig3}, not only the ground state potential but also a
so-called hybrid potential is displayed in which the gluonic
component contributes to the angular momentum.
Recently, the spectrum of such potentials has been accurately
determined by Juge, Kuti and Morningstar~\cite{colin}.
The presence of gluons in bound states should also affect
light meson and baryon spectra: one would expect
additional excitations that cannot be classified in accord with the
na\"\i{}ve constituent quark model. On the Lattice and in experiment
it should be most easy to discriminate states with exotic, i.e.\
quark model forbidden, quantum numbers
from ``standard'' hadrons.
Spin-exotic baryons cannot be constructed
but only mesons and glueballs.
First results on light hybrid mesons have been reported by
two groups~\cite{lacock}. 
The lightest spin-exotic particle has quantum numbers $J^{PC}=1^{-+}$
and a mass between 1.8 and 2~GeV.
Recent investigations incorporating sea quarks~\cite{lacock2}
confirm these findings. However, at present all experimental candidates
have masses smaller than 1.6~GeV.
Therefore, in the interpretation of experiment mixing between
spin exotic mesons and four-quark molecules, such as a $\pi f_1$,
should be considered.
It is certainly worthwhile to investigate this possibility on the Lattice
too.

\begin{figure}
\vskip -.5truecm
\epsfxsize=10truecm
\centerline{\epsfbox{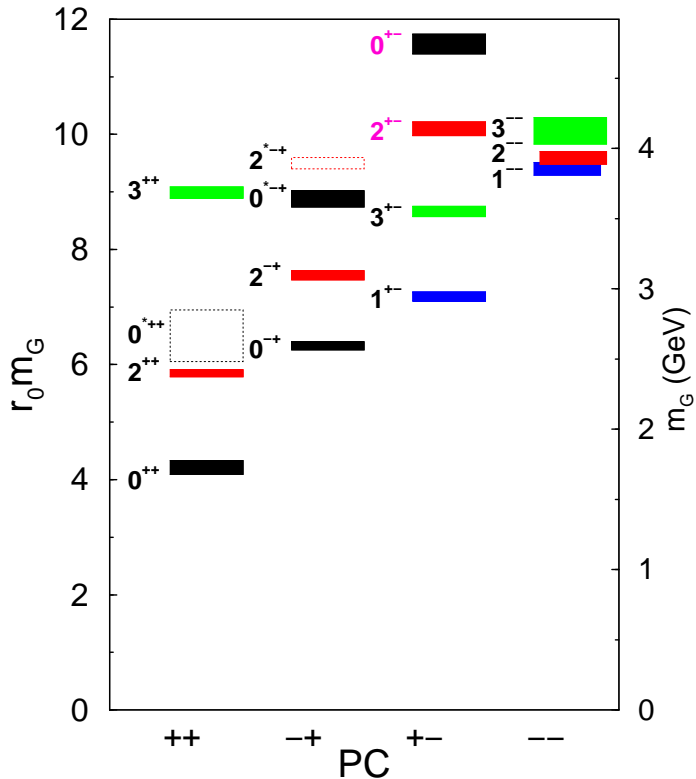}}
\vskip -.5truecm
\caption{The quenched glueballs spectrum (from Ref.~\cite{peardon}).} 
\label{fig4}
\end{figure}
Recent results 
by Morningstar and Peardon~\cite{peardon} have
revolutionised our knowledge on the quenched glueball spectrum.
Only in case of the scalar glueball
they fail to reach the precision of the 1993 state-of-the-art Lattice
predictions~\cite{baliglue}: finer lattices
are required for a safer continuum limit extrapolation.
As can be
seen from Fig.~\ref{fig4},
the ordering of glueball states has become fairly well established.
The fact that the lightest spin-exotic state $2^{+-}$ lies well above 4~GeV
explains why such states have escaped observation so far.
Glueballs are quite heavy and spatially
rather extended, due to the lack of quarks
that tie the flux together.
Therefore, these states lend themselves to the use
of anisotropic lattices: the size of the temporal lattice
spacing is dictated by the heavy mass
that one wishes to resolve while a much coarser spatial spacing
can be adapted to
resolve the glueball wave function. Introducing this anisotropy
was vital for the improvement achieved.
Recent results in QCD
with sea quarks on the scalar and tensor glueballs are
consistent with quenched findings~\cite{balifull,tcl}.
Beyond the quenched approximation, glueballs will mix with standard quark
model states. Investigations of such mixing and decay rates of
the mixed states are challenging questions~\cite{gluemix} that
are waiting to be approached
by Lattice studies in the near future.

\section{Light hadrons}
In addition to quantities that theorists or experimentalists are
interested in, well-known observables 
can be computed on the Lattice too. The motivation
is two-fold: testing QCD and gauging the Lattice methodology.
Experimental low energy input like the hadron spectrum is required
in the first place to fix the lattice spacing and quark masses.
Subsequently, among other predictions, the fundamental
parameters $\alpha_s$ and quark masses~\cite{alpha}
can be converted to, for instance, the $\overline{MS}$ scheme
that is convenient for perturbative continuum calculations.
It is not {\em a priori} clear whether the low energy
results are compatible with
values required to explain high energy QCD phenomenology.

Assuming that QCD is the right
theory, the observed states can serve as a guideline to judge
the viability of approximations, such as quenching in the absence
of high precision full QCD results.
Last but not least, quark masses and other parameters can be varied and
Lattice results can be confronted with predictions of, for instance,
chiral
perturbation theory. Indeed, evidence for quenched chiral logarithms has been
reported~\cite{cppacs}.

\begin{figure}
\vskip -.5truecm
\epsfxsize=11truecm
\centerline{\epsfbox{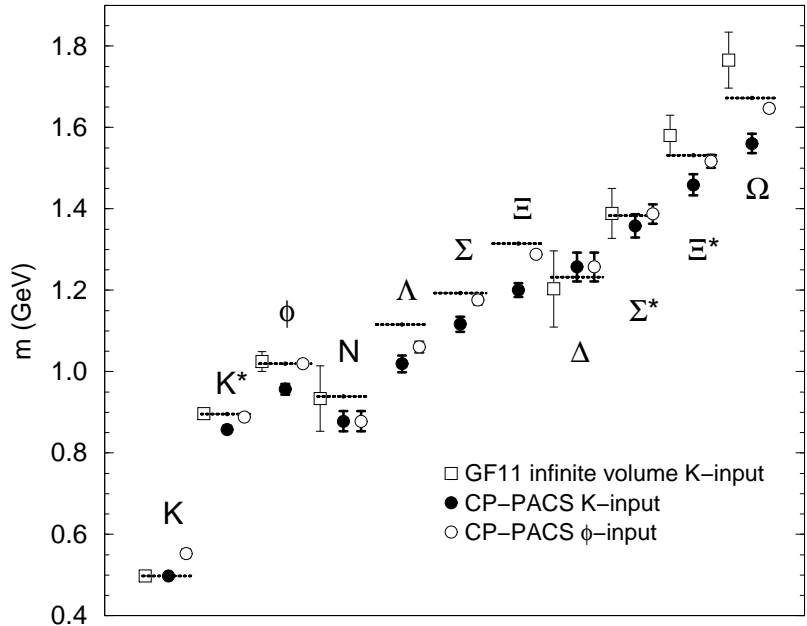}}
\vskip -.5truecm
\caption{The quenched light hadron spectrum (from Ref.~\cite{cppacs}).} 
\label{fig6}
\end{figure}
In Fig.~\ref{fig6}, I display results from a recent state-of-the-art
calculation of the quenched light hadron spectrum by the CP-PACS
collaboration~\cite{cppacs}. For comparison the results from
the GF11 collaboration~\cite{gf11hadron} that have set the standard back
in 1994 are included (squares). The $\pi$ and
$\rho$ masses that have been used as input values for the lattice
spacing $a$ and quark mass $m_u=m_d$ are omitted from the plot.
$m_s$ has been set by two methods: forcing the
$K$ mass to agree with experiment (full circles) and
forcing the $\phi$ mass to agree (open circles). Neither of the methods
can bring the spectrum completely in line with experiment.
However, no mass comes out to be wrong by more than 10~\%, indicating that
the main effect of sea quarks is to renormalise the over-all value of
the coupling, rather than altering mass ratios, despite the fact that
all particles displayed, with the exception of the nucleon,
become unstable in full QCD. First un-quenched results
by the same collaboration indicate an improvement in the direction of the
experimental values. Many groups are at present
studying quantities which one might expect to be more sensitive
towards quenching
like the $\eta'$ mass, quark masses and the $\pi N\sigma$ term~\cite{sesam}.

\section*{Acknowledgements}
I have received funding by DFG (grants Ba~1564/3-1 and Ba~1564/3-2).
I thank Andrei Afanasjev, Branko Vlahovic, Dubravko Klabucar and
Elio Soldi for organising this stimulating conference at Dubrovnik and hope
that reviving the international tradition of this
former Yugoslav science centre will serve as a
starting point to overcome nationalism, separatism and sectarian violence.

\end{document}